\documentclass[%
 reprint,
 amsmath,amssymb,
 aps,
]{revtex4-1}

\usepackage{graphicx}% Include figure files
\usepackage{dcolumn}% Align table columns on decimal point
\usepackage{bm}% bold math
\begin{document}

\preprint{APS/123-QED}

\title{Plasmon-exciton-polariton lasing}% Force line breaks with \\
%\thanks{A footnote to the article title}%

\author{Mohammad Ramezani}
\affiliation{FOM Institute DIFFER, P.O. Box 6336, 5600 HH Eindhoven, The Netherlands}%

\author{Alexei Halpin}
\affiliation{FOM Institute DIFFER, P.O. Box 6336, 5600 HH Eindhoven, The Netherlands}%

\author{Antonio I. Fern\'andez-Dom\'inguez}
\affiliation{Departmento de F\'isica Te\'orica de la Materia Condensada and Condensed Matter Physics Center (IFIMAC), Universidad Aut\'onoma de Madrid, E-28049 Madrid, Spain}%

\author{Johannes Feist}
\affiliation{Departmento de F\'isica Te\'orica de la Materia Condensada and Condensed Matter Physics Center (IFIMAC), Universidad Aut\'onoma de Madrid, E-28049 Madrid, Spain}%

\author{Said Rahimzadeh-Kalaleh Rodriguez}
\affiliation{Centre de Nanosciences et de Nanotechnologies, CNRS, Univ. Paris-Sud, Université Paris-Saclay, C2N - Marcoussis, 91460 Marcoussis, France}%

\author{Francisco J. Garcia-Vidal}
\affiliation{Departmento de F\'isica Te\'orica de la Materia Condensada and Condensed Matter Physics Center (IFIMAC), Universidad Aut\'onoma de Madrid, E-28049 Madrid, Spain}%
\affiliation{Donostia International Physics Center (DIPC), E-20018 Donostia/San Sebastian, Spain}%
 \email{fj.garcia@uam.es}

\author{Jaime G\'omez Rivas}
\affiliation{FOM Institute DIFFER, P.O. Box 6336, 5600 HH Eindhoven, The Netherlands}%
\affiliation{Department of Applied Physics, Eindhoven University of Technology, P.O. Box 513, 5600 MB Eindhoven, The Netherlands}%
 \email{j.gomezrivas@differ.nl}

%\date{\today}% It is always \today, today,
             %  but any date may be explicitly specified

\begin{abstract}
Metallic nanostructures provide a toolkit for the generation of coherent light below the diffraction limit. Plasmonicbased
lasing relies on the population inversion of emitters (such as organic fluorophores) along with feedback
provided by plasmonic resonances. In this regime, known as weak light–matter coupling, the radiative characteristics
of the system can be described by the Purcell effect. Strong light–matter coupling between the molecular excitons and
electromagnetic field generated by the plasmonic structures leads to the formation of hybrid quasi-particles known as
plasmon-exciton-polaritons (PEPs). Due to the bosonic character of these quasi-particles, exciton-polariton condensation
can lead to laser-like emission at much lower threshold powers than in conventional photon lasers. Here, we
observe PEP lasing through a dark plasmonic mode in an array of metallic nanoparticles with a low threshold in an
optically pumped organic system. Interestingly, the threshold power of the lasing is reduced by increasing the degree
of light–matter coupling in spite of the degradation of the quantum efficiency of the active material, highlighting the
ultrafast dynamic responsible for the lasing, i.e., stimulated scattering. These results demonstrate a unique roomtemperature
platform for exploring the physics of exciton-polaritons in an open-cavity architecture and pave the
road toward the integration of this on-chip lasing device with the current photonics and active metamaterial planar
technologies. 
%\begin{description}
%\item[Usage]
%Secondary publications and information retrieval purposes.
%\item[PACS numbers]
%May be entered using the \verb+\pacs{#1}+ command.
%\item[Structure]
%You may use the \texttt{description} environment to structure your abstract;
%use the optional argument of the \verb+\item+ command to give the category of each item. 
%\end{description}
\end{abstract}

%\pacs{Valid PACS appear here}% PACS, the Physics and Astronomy
                             % Classification Scheme.
%\keywords{Suggested keywords}%Use showkeys class option if keyword
                              %display desired
\maketitle

%\tableofcontents

\section{INTRODUCTION}
Exciton-polaritons—hybrid light–matter quasi-particles formed
by strong exciton-photon coupling—have inspired more than
two decades of highly interdisciplinary research ~\cite{PhysRevLett.69.3314}. Polariton
physics has largely focused on semiconductor microcavities,
where the strong nonlinearities associated with quantum well excitons \cite{PhysRevB.62.R16247,Saba:2001fj} combined with the high quality factor cavities available
through state-of-the-art epitaxial techniques have enabled the
first observations of Bose–Einstein condensation (BEC) \cite{Kasprzak:2006jy} and superfluidity \cite{Amo:2009bl} in optics.  A key early vision of the field focused
around the possibility of achieving a coherent light source at low
threshold powers without the need for population inversion: a
polariton laser~\cite{Imamoglu:1996un}. Polariton lasers have remained in the realm of proof-of-principle experiments \cite{PhysRevLett.98.126405,Sanvitto2016} and their widespread usage has not yet been adopted as efforts continue to lower thresholds and optimize operating parameters. In an effort to overcome some of the material-related limitations hampering applications of exciton-polaritons, as well as to explore novel light-matter states associated with distinct types of excitons, several researchers have recently turned their attention to organic materials~\cite{Lidzey:1998wb,Bellessa:2004fa,Dintinger:2005hfa,KenaCohen:2008hj,AberraGuebrou:2012jd}. While organic systems are generally disordered, their optical transitions can have large transition dipole moments allowing them to couple strongly to light at room temperature. Several independent studies have already reported polariton lasing/BEC \cite{KenaCohen:2010cy,Plumhof:2013bn} and nonlinear interactions with organic excitons \cite{Daskalakis:2014ex} in microcavities.

Recently, plasmonic systems have emerged as a promising
alternative platform for exploring exciton-polaritons in an open
architecture. The “cavity” defining the resonator is no longer a
multilayer dielectric stack possessing a complex spectral response,
and facilitates the integration of exciton-polariton devices with
integrated photonics circuits. In these plasmonic systems, previously
shown to be highly suitable for photon lasing \cite{Oulton2009,PhysRevLett.110.206802,Suh:2012go,Zhou:2013ef,Schokker:2014bw}, the
excitonic material can be easily integrated by solution processing.
The quality factors of plasmonic resonances are much lower
than their counterparts in dielectric microcavities. However, the
subwavelength field enhancements generated by resonant metallic
nanostructures can significantly boost the light–matter coupling.
Indeed, strong plasmon-exciton coupling has already been
observed ~\cite{Dintinger:2005hfa,PhysRevLett.103.053602,AberraGuebrou:2012jd,Rodriguez:2013hi,Rodriguez:OX,Vakevainen:2014,Eizner:2015it,Zengin:2015kv}, but earlier attempts toward achieving plasmon-exciton-polariton (PEP) lasing remained unsuccessful due to the inefficient relaxation mechanism of PEPs and the saturation of strong coupling at large pumping fluences \cite{Rodriguez:2013hi}.  
Here, we demonstrate PEP lasing from an optically pumped array of silver
nanoparticles coated by a thin layer of organic molecules at room
temperature, occurring at a low threshold ~\cite{Sanvitto2016}. Strong coupling between
excitons in the organic molecules and collective plasmonic
resonances of the array forms PEPs. By increasing the PEP density
through optical pumping, we observe a pronounced threshold in
the emission intensity accompanied by spectral narrowing. Besides
these generic lasing characteristics, our system exhibits two rather
distinct features: first, the threshold power for PEP lasing is reduced
in parallel with a degradation of the quantum efficiency
of the material. This counterintuitive behavior, from the standpoint
of conventional laser physics, is intimately related to the
onset of strong coupling and the emergence of new eigenstates,
i.e., PEPs. A second distinct feature of our PEP laser stems from
the fact that the nanoparticle array supports dark as well as bright
modes. The mode that first reaches the lasing threshold is in
fact dark below the threshold. While dark-mode photon lasing
has attracted significant interest in the plasmonics community
for several years, we provide the first report of lasing from a dark
mode in a strongly coupled plasmon-exciton system. Lasing from
this dark (below threshold) mode also manifests in an abrupt
polarization rotation of the emitted light by 90$^\circ$ above threshold.

We first characterize strong coupling between excitons in
an organic dye and the lattice modes supported by an array of
silver nanoparticles through optical extinction measurements.
Subsequently, using numerical and semi-analytical techniques,
we analyze the modes supported by the array and the composition
of the associated PEPs, respectively. Photoluminescence (PL)
measurements on the sample pumped off-resonance provide
the emission response at increasing PEP densities. At high emitter
concentrations, we observe Rabi splitting, the signature of
plasmon-exciton strong coupling, together with the appearance of
stimulated scattering and PEP lasing. By measuring the dispersion
of the PL at several pump fluences for both polarizations, we
identify the dark mode responsible for PEP lasing in this system.

\section{Results and Discussion}

\begin{figure*}[tbhp]
\centering
\fbox{\includegraphics[width=1\linewidth]{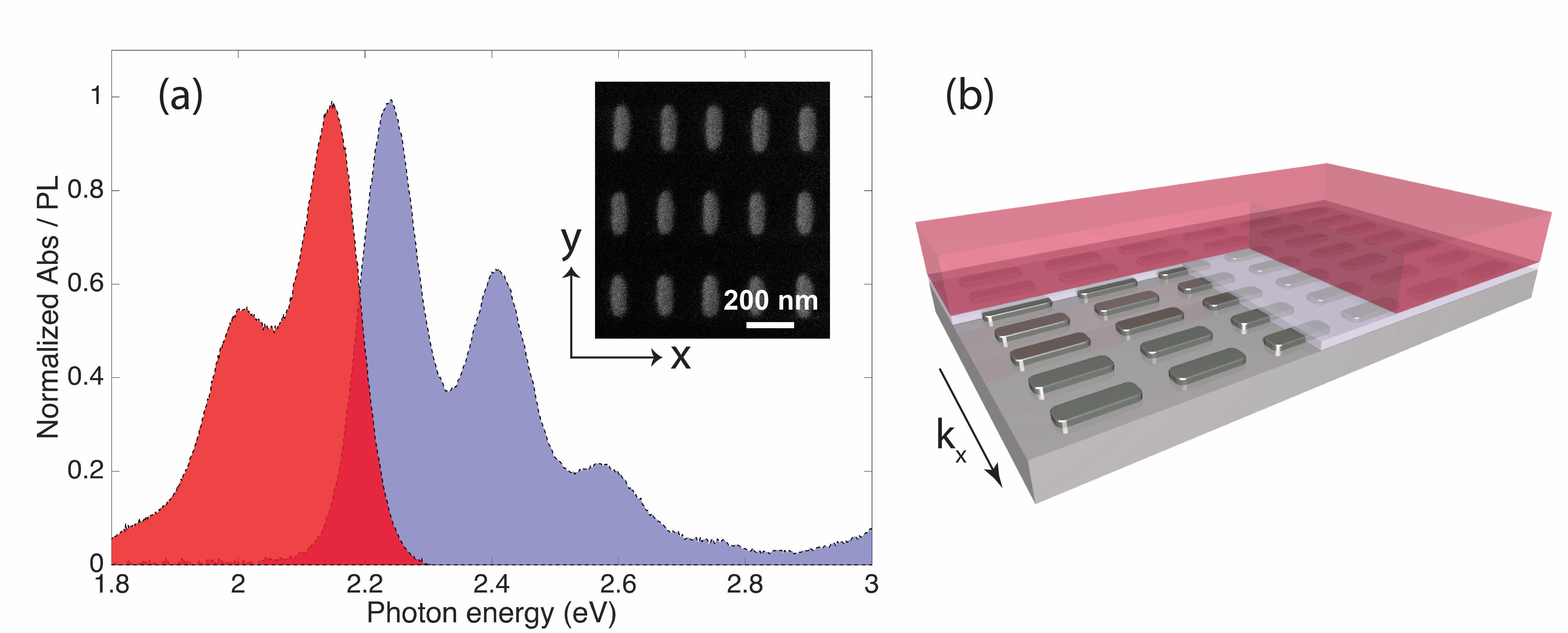}}
\caption{(a) Normalized absorption (blue) and photoluminescence (red) spectra of the layer of PMMA doped with dye molecules in the absence of the plasmonic array. The inset shows an SEM image of the array of silver nanoparticles. (b) Schematic illustration of the array covered with a thin layer of PMMA doped with dye molecules.}
\label{fig:Sample}
\end{figure*}

Fig.~\ref{fig:Sample} shows the normalized absorption and PL spectra of a layer of PMMA doped with organic dye molecules. We use a rylene dye [N,N$^\prime$-Bis(2,6-diisopropylphenyl)-1,7- and -1,6-bis(2,6-diisopropylphenoxy)-perylene-3,4:9,10-tetracarboximide] as an emitter \cite{patent}, due to its high photostability and low propensity towards aggregation at high concentrations. Two distinct peaks corresponding to the main electronic transition at E = 2.24 eV and first vibronic sideband at E = 2.41 eV are evident in the absorption spectra of the molecules. A layer of PMMA doped with dye molecules with thickness of 260 nm is spin-coated on top of the plasmonic array of silver nanoparticles.  A SEM image of the fabricated array is depicted in the inset of Fig.~\ref{fig:Sample}. The array consists of particles with dimensions of $200~nm\times70~nm\times20~nm$ and the pitch sizes along the $x$ and $y$ directions are 200 nm and 380 nm, respectively.

\begin{figure*}[tbhp]
\centering
\fbox{\includegraphics[width=1\linewidth]{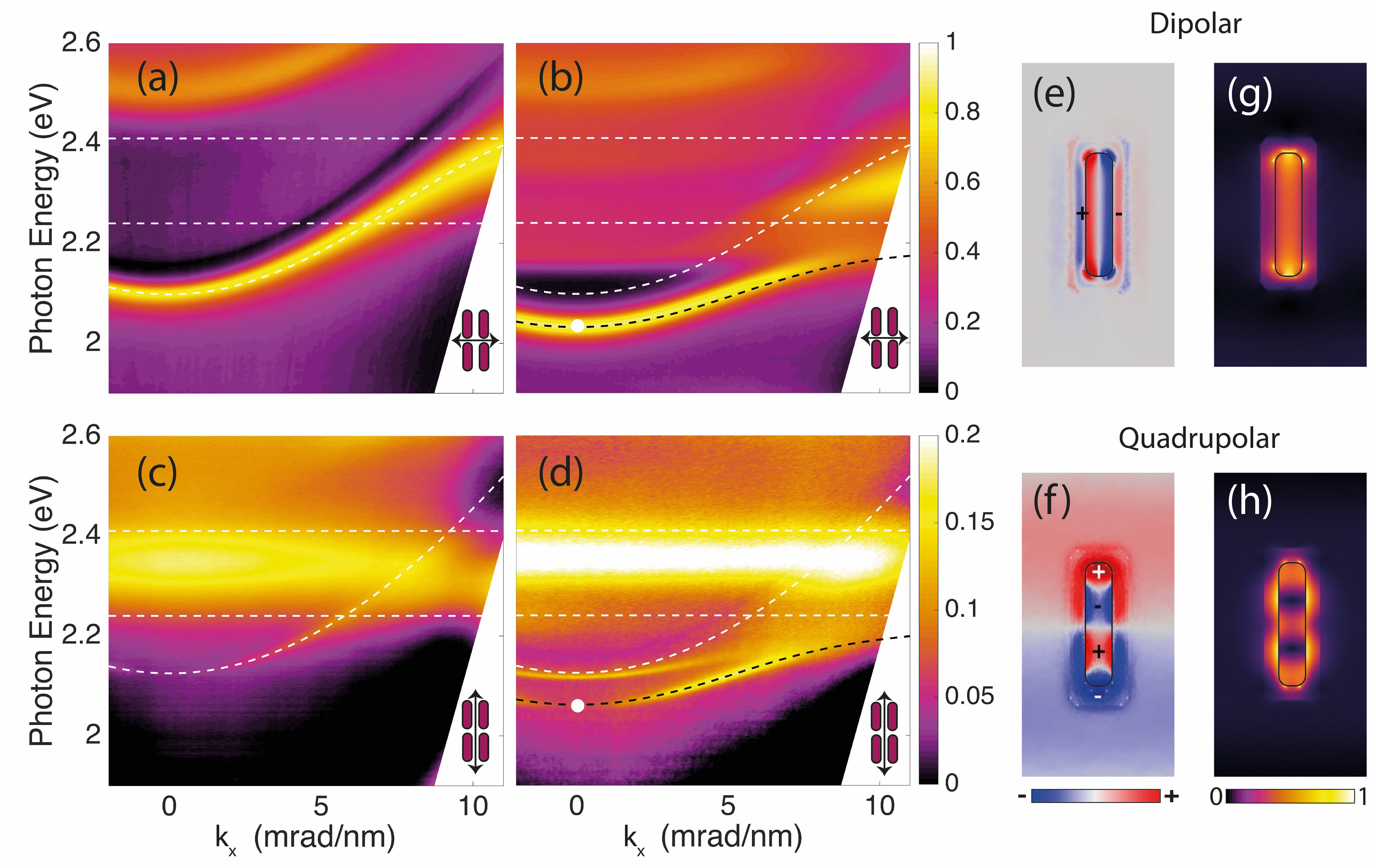}}
\caption{The extinction measurements of the sample with incident light polarized along the short axis of the nanoparticles in the absence (a) and the presence (b) of the dye molecules. The extinction of the same sample illuminated by the light polarized along the long axis of the nanoparticles without (c) and with (d) the dye. The white dashed lines indicate the energies of the two strongest vibronic transitions in the molecules (horizontal lines) and the SLR dispersion whereas the dashed black lines correspond to PEP modes. Maps of the induced charge density (e,f) and normalized electric field amplitude (g,h) for the corresponding lowest PEP modes at $k_{x}$ = 0 mrad/nm. (e,g) correspond to the bright (dipolar-like) resonance, while (f,g) to the dark (quadrupolar) resonance at $k_{x}$ = 0 mrad/nm.}
\label{fig:Extinction}
\end{figure*}

First, we measure the angle-resolved extinction of the nanoparticle array when covered by an undoped layer of PMMA (Figs.~\ref{fig:Extinction}(a,c)). The polarization of the incident light is fixed, and set to be either perpendicular (top row) or parallel (bottom row) to the long axis of the nanoparticles as indicated by the arrow in the inset. Here we take advantage of a particular type of plasmonic modes that are supported by periodic arrays of metal nanoparticles, the so-called surface lattice resonances (SLRs). These modes are the result of the radiative coupling between localized surface plasmon (LSP) resonances in the individual nanoparticles enhanced by the in-plane diffracted orders of the array, i.e., the Rayleigh Anomalies (RAs). Energy dispersions and quality factors of these SLRs can be tailored by varying the geometrical parameters and energy detuning between RAs and LSP resonances \cite{RevModPhys.79.1267,doi:10.1021/ph400072z}. In addition, the enhanced in-plane radiative coupling reduces the radiative losses associated to localized resonances \cite{:/content/aip/journal/jcp/120/23/10.1063/1.1760740} and the redistribution of the electromagnetic field around the particles also reduces Ohmic losses \cite{:/content/aip/journal/jap/118/7/10.1063/1.4928616}, creating narrow resonances with high quality factors \cite{:/content/aip/journal/jcp/120/23/10.1063/1.1760740,Humphrey:2014dd,PhysRevLett.102.146807}.

By probing the sample under different polarizations and angles of incidence, we couple to different resonances with distinct electric field distributions and symmetries depending on which RAs are probed. The resonance with parabolic dispersion in Fig.~\ref{fig:Extinction}(a) demonstrates the combination of both strong extinction and narrow linewidth in SLRs, which results in this case from the enhanced radiative coupling of (0,$\pm$1) RAs to a dipolar LSP resonance supported by the individual nanoparticles (see below). Similarly, by rotating the polarizer and aligning the incident electric field along the long axis of the nanoparticles, we probe the coupling of (0,$\pm$1) RAs to a quadrupolar LSP resonance supported by the rods (see also below) \cite{Giannini:2010ip}. Differences between the extinction measurements of Figs.~\ref{fig:Extinction}(a,c) can be observed, where the most important is the vanishing SLR at $k_{x}= 0$~mrad/nm for incident electric field parallel to the long axis of the rods. For this case the electric field distribution is not dipole-active along the polarization direction. As a result, coupling of the free space radiation into this mode is inefficient and the mode is dark at normal incidence. This behaviour implies a strong reduction of radiation losses of this mode which, as we will show later, is a key ingredient to achieve PEP lasing \cite{PhysRevB.64.125122, PhysRevB.85.245411}. Due to the multipolar character of the dark mode, the net dipole moment is not zero (Fig.~\ref{fig:Extinction}). This residual dipole moment is responsible for the out-coupling of the PL above threshold that is reported ahead. Extinction measurements of the same array in the presence of a 260 nm thick PMMA layer doped with dye molecules are shown in Figs. \ref{fig:Extinction}(b,d). The dye concentration (C) in the polymer is 35 wt$\%$ and the measurements are referenced to the extinction of the doped layer without the nanoparticles. Increasing the molecular concentration modifies the extinction dispersion as a result of the hybridization between the SLR and the molecular transition. For both polarizations we observe an anticrossing in the extinction dispersion and an associated Rabi splitting, $\hbar\Omega = 200$~meV, at $k_{x}\approx7$~mrad/nm, which reveals the emergence of strong coupling between the SLR and molecular excitations. The coupling leads to the creation of upper and lower PEPs where hybridization occurs for both bright and dark modes. In Fig. \ref{fig:Extinction}(d), another resonance slightly blue shifted with respect to the lower PEPs is visible. This resonance corresponds to the guided-mode supported by the polymer layer. The appearance of this guided mode is due to the increase of the refractive index of the doped polymer layer when the molecular concentration is increased.

In order to analyze the hybrid light-matter nature of the PEPs formed by strong coupling of the SLRs to the dye molecules, we use a few-level Hamiltonian that reproduces the measured extinction dispersions and allows the determination of the Hopfield coefficients defining the exciton and photon components of the strongly coupled system. We treat each field polarization separately, with Hamiltonian (for each in-plane momentum)

$$
H= \begin{pmatrix}
E_{SLR}&g_{1}&g_{2}\\
g_{1}&E_{dye,1}&$0$\\
g_{2}&$0$&E_{dye,2}\\
\end{pmatrix}
$$

where $E_{SLR}$ is the energy of the SLR, $E_{dye,1}$ ($E_{dye,2}$) is the dispersionless energy of the first (second) absorption peak of the dye, and $g_1$ ($g_2$) describes the coupling between the SLR and the molecular modes. In Fig. \ref{fig:Extinction}, both the input energies of the ``bare'' states and the energy of the lowest PEP modes are depicted. As the actual system contains additional states at higher energies that we do not treat, we only show the lowest coupled state in the figure. From this analysis we can extract the Hopfield coefficients of the lower PEP, indicating that at $k_{x}=0$~mrad/nm it is composed of $75\%$ SLR and $25\%$ molecular excitations, with similar values obtained for both polarizations. We note here that while the SLR for the polarization along the long axis of the particles is dark under far-field excitation at $k_{x}=0$~mrad/nm, the coupling to the molecules occurs in the near-field and thus attains similar strengths for both polarizations. 

Electromagnetic calculations were also performed to simulate
the extinction properties of the experimental samples. Both the
finite-difference time-domain (FDTD) (Lumerical) and finite
element method (FEM) (Comsol Multiphysics) methods were
employed, finding an excellent agreement with the extinction maps in Fig. \ref{fig:Extinction} (see Supplementary Information for details). Geometric and material parameters were extracted from SEM images of the samples and previous literature \cite{palik}, respectively. The right insets in Fig. \ref{fig:Extinction} render electric field amplitude and induced surface charge density maps evaluated at $k_{x}$ = 0~mrad/nm. The top insets (Fig. \ref{fig:Extinction}(e,g)) show that, for a polarization along the short axis, the near-field is governed by the bright dipolar-like LSP supported by the rods, i.e., the so called ($\lambda/2$) LSP, as anticipated above. On the contrary, the bottom insets (Fig. \ref{fig:Extinction}(f,h)) indicate that a dark quadrupolar LSP, i.e., the ($3\lambda/2$) LSP, resonates for incoming light polarized along the long axis of the rod. Importantly for lasing purposes, both polaritonic modes spectrally overlap within an energy window of a few meV.    
The vacuum Rabi frequency $\Omega$ defining the coupling strength is proportional to $\sqrt{N}$, where $N$ is the number of excitons within the mode volume of the resonance. Therefore, achieving strong coupling with organic molecules requires increasing their concentration within the polymer matrix such that the exciton density is maximized \cite{Rodriguez:OX,Vakevainen:2014,Suh:2012go}. However, one immediate drawback expected from increasing C for the emission is the emergence of considerable inter-molecular interactions, which can lead to spontaneous aggregation (modifying the spectral properties of the sample) and also to a reduction of fluorescence lifetime ($\tau$) due to the enhancement of non-radiative decay channels (concentration quenching of the emission) \cite{Lakowicz}. Importantly, the dye molecules used here do not suffer from aggregation at high C as the normalized absorption spectrum for different C remains unchanged (See Fig. S4 at the Supplementary Information for details). To quantify the concentration quenching, we have measured the lifetime and quantum efficiency of the polymer layers at different C (see Supplementary Information for details). One can see that at low C the quantum efficiency of the molecules is close to the unity and the lifetime of the excited molecules is 5.8 ns. However, by increasing C to 35 wt$\%$ the quantum efficiency is reduced to 3 $\%$ with a corresponding $\tau= 400$ ps. The emissive properties of high C samples are thus unsuitable for stimulated emission and photon lasing, where a high quantum yield is desired for achieving gain. In fact, the prediction that lasing without inversion could be achieved by exploiting many-body coherence in strongly coupled exciton-photon systems has been a major motivation for the development of exciton-polariton lasers  \cite{Imamoglu:1996un}.

\begin{figure*}[tbhp]
\centering
\fbox{\includegraphics[width=1.05\linewidth]{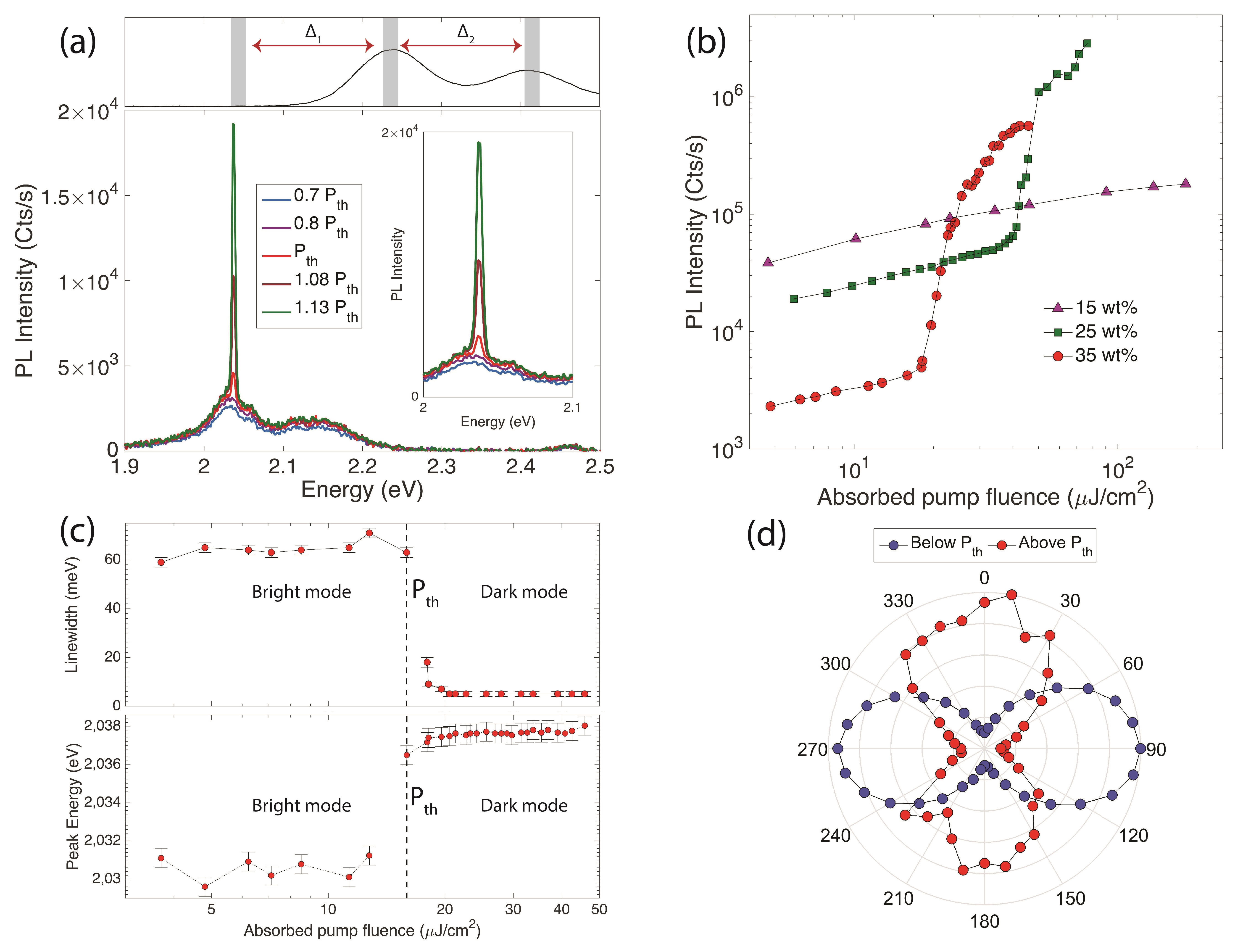}}
\caption{(a) Emission spectra along the forward direction for the array of nanoparticles covered with PMMA with the dye concentration of 35 wt $\%$ at increasing absorbed pump fluences. (Inset) Close view of the lasing peak. Upper panel: Absorption spectrum of the dye (solid curve). The energies of the lasing emission, the main electronic transition and the first vibronic side band of the molecule are indicated by the gray shaded areas. Note that the energy differences $\Delta_{1}$ and $\Delta_{2}$ are equal. (b) Photoluminescence peak intensity as a function of absorbed pump fluence for three samples at different dye concentrations. (c) Linewidth and energy shift of the photoluminescence peak as a function of absorbed pump fluence for the sample with C = 35 wt$\% $ dye concentration. The linewidths and peak energies are extracted by fitting a Gaussian function to the spectrum. The errorbars in peak energy plot are set by the resolution of the spectrometer ($\approx$ 1 meV) (d) Polarization of the emission from the sample with C = 35 wt$\%$ below and above threshold (P = 1.5P$_{th}$). The long axis of the nanoparticles is oriented along the vertical direction ($\theta= 0^{\circ}$).}
\label{fig:Powerscan}
\end{figure*}

In order to investigate PEP lasing, we optically pump the sample nonresonantly and measure the PL spectra as a function of the incident excitation power for different concentration of the molecules. The wavelength of the pump laser is centered at $\lambda_{exc}= 500$~nm ($E_{exc} = 2.48$~eV) and the pump polarization is fixed along the short axis of the nanoparticles for all experiments. The spectra of the PL for the sample with C = 35 wt$\%$ in the forward direction for different absorbed pump fluences is shown in Fig.~\ref{fig:Powerscan}(a), where we observe the appearance of a very sharp peak in the emission spectrum at a fluence of $18~\mu$J/cm$^{2}$. In Fig.~\ref{fig:Powerscan}(b), we plot the maximum of the PL intensity as a function of absorbed pump fluence for three different concentrations of dye. For the low concentration sample, C = 15 wt$\%$, the PL intensity increases linearly with the incident power. This sample is at the onset of the appearance of exciton-SLR hybridization in extinction (see Supplementary Information).  At C = 25 wt$\%$, we observe the emergence of a threshold in the PL, followed by a superlinear increase of the emission. The nonlinear response of the system is further increased at C = 35 wt$\%$ where the threshold fluence is lowered by a factor of 2. This threshold fluence is one of the lowest values observed in optically pumped organic polariton lasers~\cite{KenaCohen:2010cy,Plumhof:2013bn,Daskalakis:2014ex} and also is lower than the reported threshold for plasmonic based photon lasers in the weak coupling regime~\cite{Zhou:2013ef,Sidiropoulos2014}. Note that in our experiments, photo-degeradation occurs before the transition to the weak coupling regime ~\cite{Rodriguez:2013hi}, so the photon lasing is precluded by the damage threshold. Moreover, it is interesting to note that
below the threshold, the emission decreases as C increases. This
decrease is mostly due to the reduction of the photoluminescence
quantum yield (PLQE) (See Supplementary Information).

The linewidth of the PL as a function of the absorbed pump fluence is shown in Fig.~\ref{fig:Powerscan}(c), where a strong reduction of the linewidth above threshold and hence, substantial increase of the temporal coherence, can be observed. In Fig.~\ref{fig:Powerscan}(d) we display the polarization of the emission below and above the threshold for the sample with C = 35 wt$\%$. Below the threshold, the emission of the sample is mainly polarized along the short axis of the particles, i.e., 90$^{\circ}$. This emission is associated with PEPs originating from the hybridization of the bright SLR with excitons (see Fig.~\ref{fig:Extinction}(b)). Interestingly, above the threshold, the polarization of the emission rotates by 90$^{\circ}$ and is primarily oriented along the long axis of the particles. This polarization rotation strongly indicates that the emission above threshold is dominated by the PEPs created from the dark modes whose polarization point to the long axis of the particles (see Fig.~\ref{fig:Extinction}(d)). 

\begin{figure*}
\centering
\fbox{\includegraphics[width=1.05\linewidth]{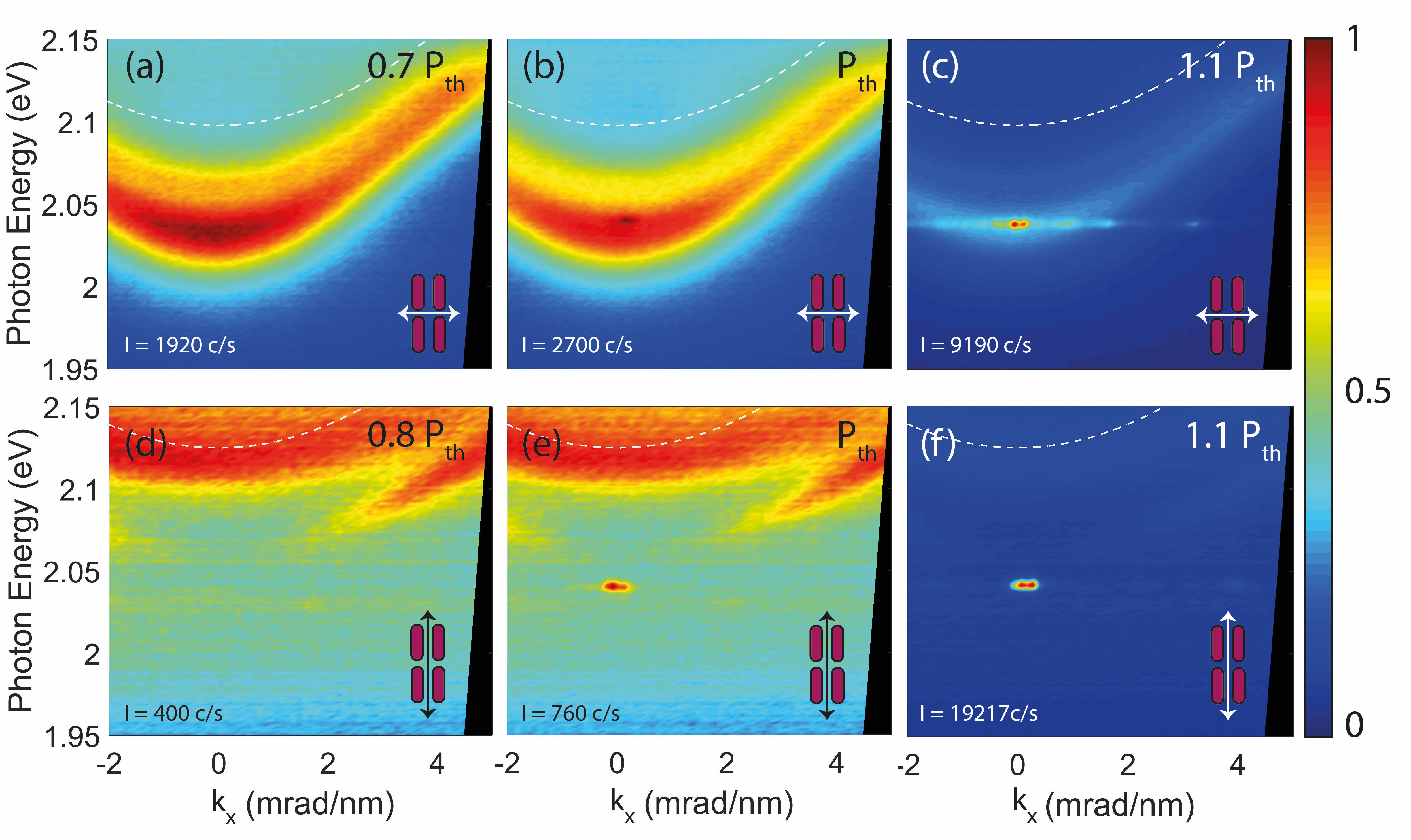}}
\caption{Normalized angular-resolved emission measurements for two detection polarizations, parallel and orthogonal to the long axis of the nanoparticles, for the array of nanoparticles covered with PMMA with the dye concentration of 35 wt $\%$. The cartoon in the inset of each panel depicts the orientation of the nanoparticles with an arrow indicating the direction of the transmission axis of the polarization analyzer. The emission intensity in unit of counts/s at E = 2.04 eV and k$_{x}=0$ mrad/nm for each detection condition is indicated at the bottom of each panel. (a-c) Angle-resolved PL for different pump fluences with the analyzer along the short axis of the nanoparticle visualizing to the bright mode along (0,$\pm$1) RAs supported by the array. (d-f) Angle-resolved PL for different pump fluences with analyzer along the long axis of the nanoparticle. In this configuration, the dark mode excited by (0,$\pm$1) RAs is probed. The bare modes associated with the uncoupled SLRs are shown by white dashed line in each panel.}
\label{fig:anglePL}
\end{figure*}

After excitation the molecules relax to the PEP band. In this incoherent relaxation, the in-plane momentum is not conserved, leading to a broad distribution of excitation across the whole band. Most of the excitation thus ends up in high-$k_{x}$ reservoir modes that are almost uncoupled from the SLR \cite{Kasprzak:2006jy}. In order to achieve polariton lasing, the population of a single state has to become large enough to obtain significant bosonic stimulation. In semiconductor microcavities, the necessary relaxation from the reservoir typically relies on exciton-polariton and polariton-polariton scattering. In contrast, microscopic models for organic polariton lasing have suggested that vibronic coupling can play a more important role for dissipating excess momentum than exciton-polariton and polariton-polariton scattering \cite{PhysRevB.80.235314,Mazza:2013dr, Cwik:2014ex}. In this picture, high frequency intramolecular vibrational modes allow exciton-polaritons to scatter directly from the reservoir towards the lower energy levels  \cite{Somaschi:2011fd}. This implies that organic exciton-polariton lasing is most efficient when the relaxation from the reservoir to the lasing state is resonant with a strong optical phonon line \cite{KenaCohen:2010cy}. As shown in the top inset of Fig.~\ref{fig:Powerscan}(a), this condition is exactly fulfilled in the current sample: the energy difference, $\Delta= \Delta_{1}= \Delta_{2}$, between the vibronic subpeaks of the molecule (the phonon energy) corresponds exactly to the energy difference between the lowest peak and the lasing state. Moreover, one needs to take into account that in addition to the role of the lattice parameters in modifying the detuning between the SLR and the molecular exciton, the change of the molecular concentration can also alter the detuning through the change in the refractive index of the layer. The specific energy at which the polariton lasing occurs also poses an interesting question: while energy shifts are seen as the smoking-gun for interactions between polaritons in semiconductors microcavities, the reported spectral behaviour of polariton lasers in organic systems has been variable \cite{KenaCohen:2010cy,Daskalakis:2014ex, Plumhof:2013bn}. In our system as the PL  below and above threshold is dominated by different modes, we must distinguish the values of the energy shift for these two regimes. In Fig.~\ref{fig:Powerscan}(c), one can see that above the threshold, the dominant photoluminescence originating from the dark mode shifts by 1.3 meV and locks at 2.038 eV. Locking of the energy shift above threshold has been predicted by the model in Ref. \cite{Daskalakis:2014ex}. This is in agreement with our measurements considering organic polariton interaction within the condensate. 
Similar to the extinction measurements shown earlier, we can use angular-resolved measurements to study PEP emission at increasing pump fluences as we approach the critical density of PEPs \cite{Rodriguez:2013hi}. In Fig.~\ref{fig:anglePL}, we display the measured dispersion of the emission both below and above threshold for two orthogonal detection polarizations corresponding to the bright and dark modes seen in Figs.~\ref{fig:Extinction}(b,d). In Figs.~\ref{fig:anglePL}(a,d), where the pump fluence is below threshold, we recover the same dispersions as those shown previously in the extinction measurements, with a bright mode for horizontal polarization and a dark mode for vertical. We observe emission over the whole range of $k_{x}$ indicating that, as mentioned above, the molecular relaxation process populating the PEPs after high-frequency excitation does not conserve $k_{x}= 0$~mrad/nm. Moreover, the PL from the uncoupled molecules lead to the green and cyan background in Figs.~\ref{fig:anglePL}(a,d). Upon increasing the pump fluence, we observe a collapse in the emission pattern towards $k_{x} = 0$~mrad/nm over the narrow spectral range seen in the spectra of Fig.~\ref{fig:Powerscan}(a). As we mentioned earlier, while the system exhibited no vertically polarized emission at normal incidence below threshold, above threshold the lasing peak is mainly polarized along this direction. This behaviour highlights one of the major differences between open systems defined by plasmonic lattices and traditional microcavities: the bright mode corresponding to the SLR of Fig.~\ref{fig:Extinction}(a) represents a lossy mode due to the radiation losses as it can efficiently couple out into free-space. On the other hand, the dark mode is significantly less lossy at $k_{x}=0$~mrad/nm due to suppression of radiation losses, which favors PEP lasing at this mode. The PEPs created via this mode can be accumulated with a much lower probability of decay. The energy dispersion of the bare modes associated with uncoupled SLRs for the dark and bright modes are shown in Fig.~\ref{fig:anglePL} indicating that the system remains in the strong coupling regime. We also observe in Fig.~\ref{fig:anglePL}(c),a residual PL emission with a flat dispersion at the energy in which the lasing occurs. This emission is due to the scattering and polarization conversion of the PEP lasing emission from local imperfections in the sample.

\section*{Conclusion}

In conclusion, we demonstrate the first polariton laser based on a plasmonic open cavity using an organic emitter. By exploiting diffractive coupling in a metallic nanoparticle array, we obtain spectrally sharp surface lattice resonance modes leading to the formation of PEPs in a dye-doped layer above the surface of the array. A lasing threshold in this system appears alongside the onset of strong-coupling at high dye concentrations, when a sufficiently large number of molecules interact with the SLR mode, leading to an avoided crossing as seen in extinction measurements. At subsequently higher concentrations, the threshold pump energy diminishes and we observe a very low lasing threshold in organic polariton laser systems. Though the observation of strong-coupling appears in far-field measurements through coupling of a bright mode to molecular resonances, the radiatively dark mode supported by the array plays a critical role in the near-field providing a low-loss channel in which polaritons can accumulate. This ultimately results in PEP lasing at an orthogonal polarization to that of the bright mode. While the optimal conditions for polariton lasing in plasmonic arrays remain to be determined, subsequent tailoring of both the mode structure of the array and the emitter could conceivably lower the lasing threshold reported here further (for instance by making use of pump-enhancement). Note that the energy difference between the RAs and the LSP resonance in arrays of nanoparticles defines the confinement of SLRs to the surface, i.e., the mode volume and also their quality factor which affect the threshold. Therefore, arrays of metallic nanoparticles constitute a rich system that requires a through investigation. In addition, the ease in fabrication of large area arrays by nanoimprint lithography and the open nature of the cavity, opens a window through other interesting phenomena occurring in polariton systems with an architecture that provides potentially straightforward implementation into devices.

\section*{Methods}

\subsection*{Fabrication}

The array of silver nanoparticles was fabricated by substrate conformal nanoimprint lithography following the procedure described in Ref.\cite{Marc} on glass (Eagle 2000) and encapsulated by 8 nm of SiO$_{2}$ and 20 nm of Si$_{3}$N$_{4}$ to prevent oxidation of silver.

\subsection*{Characterization}

The extinction measurements have been performed using a collimated polarized white light beam generated from a halogen lamp. The zeroth-order transmission of the white light from the sample under different incident angles ($T_{0}(\theta)$) is collected. For the reference measurement, the transmitted light through the same substrate and the polymer layer in the absence of the array ($T^{ref}_{0}(\theta)$) is measured. The extinction is defined as $1- T_{0}(\theta)/T^{ref}_{0}(\theta)$.

All photoluminescence measurements except for the polarization measurement below threshold in Fig. 3(b) have been done with amplified pulses generated from an optical parametric amplifier with an approximate pulse duration of 100 fs and a repetition rate of 1 kHz. Excitation using a low repetition rate but high pulse energies allows to generate large densities of PEPs without the thermal damage associated to high duty-cycle excitation. The excitation laser is focused on 100 $\mu$m diameter spot on the sample. The polarization measurements below threshold were done with a continuous-wave diode laser emitting at $\lambda$ = 532 nm.

\section*{Funding Information}
This research was financially supported by the Nederlandse Organisatie voor Wetenschappelijk Onderzoek (NWO) through the project LEDMAP of the Technology Foundation STW and through the Industrial Partnership Program Nanophotonics for Solid State Lighting between Philips and the Foundation for Fundamental Research on Matter FOM. This work has been also funded by the European Research Council (ERC-2011-AdG proposal No. 290981), by the European Union Seventh Framework Programme under Grant Agreements  FP7-PEOPLE-2013-CIG-630996 and FP7-PEOPLE-2013-CIG-618229, and the Spanish MINECO under Contract No.~MAT2014-53432-C5-5-R.

\section*{Acknowledgments}

We sincerely thank Martin K{\"o}nemann for providing us the organic dye. We are grateful of Marc A. Verschuuren for the fabrication of the samples.

\section*{Supplemental Documents}
See Supplementary document for supporting content.

%\bibliography{References}

\pagebreak
\newpage 

\newpage 

\newpage

\newpage 

\newpage

\end{document}